\documentclass[aps,prl,reprint,groupedaddress,longbibliography,superscriptaddress]{revtex4-2}
\usepackage{placeins}
\usepackage{amsmath,amsfonts,amssymb}                        
\usepackage{amsmath}
\usepackage{graphicx}
\usepackage[colorlinks=true, allcolors=blue]{hyperref}
\usepackage{siunitx}
\usepackage{changes}
\usepackage{bm}
\usepackage{float}
\usepackage{hyperref}
\usepackage{lmodern}
\graphicspath{{../Figures/PDF/}}
\definechangesauthor[name={Dmitry Bykov}, color=red]{DB}
\definechangesauthor[name={Tracy Northup}, color=purple]{TN}

\begin{document}

\title{3D sympathetic cooling and detection of levitated nanoparticles}

\author{Dmitry S. Bykov}
\email[]{dmitry.bykov@uibk.ac.at}
\address{Institut f{\"u}r Experimentalphysik, Universit{\"a}t Innsbruck, Technikerstra\ss e 25, 6020 Innsbruck,
	Austria}
\author{Lorenzo Dania}
\address{Institut f{\"u}r Experimentalphysik, Universit{\"a}t Innsbruck, Technikerstra\ss e 25, 6020 Innsbruck,
	Austria}
\author{Florian Goschin}
\address{Institut f{\"u}r Experimentalphysik, Universit{\"a}t Innsbruck, Technikerstra\ss e 25, 6020 Innsbruck,
	Austria}
\author{Tracy E. Northup}
\address{Institut f{\"u}r Experimentalphysik, Universit{\"a}t Innsbruck, Technikerstra\ss e 25, 6020 Innsbruck,
	Austria}
\date{\today}

\begin{abstract}
Cooling the center-of-mass motion of levitated nanoparticles provides a route to quantum experiments at mesoscopic scales.
Here we demonstrate three-dimensional sympathetic cooling and detection of the center-of-mass motion of a levitated silica nanoparticle. The nanoparticle is electrostatically coupled to a feedback-cooled particle while both particles are trapped in the same Paul trap. We identify two regimes, based on the strength of the cooling: in the first regime, the sympathetically cooled particle thermalizes with the directly cooled one, while in the second regime, the sympathetically cooled particle reaches a minimum temperature. This result provides a route to efficiently cool and detect particles that cannot be illuminated with strong laser light, such as absorptive particles, and paves the way for controlling the motion of arrays of several trapped nanoparticles.

\end{abstract}

\maketitle
Levitated particles are a promising experimental platform for testing fundamental physics and 
detecting weak forces~\cite{neukirch2015nanooptomechanics,millen2020optomechanics,gonzalezballestero2021levitodynamics}. 
Having recently entered the quantum regime~\cite{delic2020cooling,tebbenjohanns2020motional,tebbenjohanns2021quantum,magrini2021realtime}, levitated particles offer a means to study 
quantum mechanics at mesoscopic scales
and to build sensors with quantum-enhanced sensitivity. In particular, it has been proposed to use levitated particles with internal degrees of freedom, such as magnetic particles or nanodiamonds with color centers, to create macroscopic superposition states~\cite{yin2013large, scala2013matterwave} and to study spin-mechanical coupling~\cite{wachter2021optical}. However, optical trapping of such absorptive particles in high and ultra-high vacuum is challenging due to the increase in their bulk temperature, which can lead to burning and graphitization~\cite{neukirch2015multidimensional,delord2017diamonds,rahman2016burning}. Moreover, in the quantum regime, elevated internal temperature is associated with the decoherence of the center-of-mass (CoM) motion~\cite{kaltenbaek2016macroscopic}. Electrodynamic levitation removes the trapping laser field as a heating mechanism and has already been employed to trap absorptive particles~\cite{delord2017diamonds,Conangla2018,KuhlickeSchellZollEtAl2014,nagornykh2017optical}. 
Nonetheless, state-of-the-art experiments with electrodynamic traps still rely on laser fields to cool~\cite{dania2021optical,millen2015cavity} and interrogate the particle~\cite{Conangla2018,nagornykh2015cooling}.

Sympathetic methods provide a solution: if an absorptive particle is electrostatically coupled to one or more transparent particles, 
cooling and detection of the absorptive particle can be achieved when only the transparent particles are exposed to laser light; the absorptive particle remains in the dark.
Sympathetic cooling has been realized in experiments with neutral atoms~\cite{myatt1997production}, atomic ions~\cite{bruzewicz2019trappedion}, molecular ions~\cite{willitsch2012coulombcrystallised}, positrons~\cite{baker2021sympathetic} and proteins~\cite{offenberg2008translational}, while sympathetic detection has made possible indirect observation of ions in a Paul trap~\cite{willitsch2012coulombcrystallised,guggemos2015sympathetic}. Recently, one-dimensional sympathetic cooling has been demonstrated with two levitated micro- and nanoparticles coupled via electrostatic~\cite{slezak2019microsphere,penny2023sympathetic} or light forces~\cite{arita2022alloptical}. However, stabilization of the motion in the two remaining dimensions is still necessary to counteract trapping instabilities~\cite{delic2020cooling,tebbenjohanns2020motional,tebbenjohanns2021quantum,magrini2021realtime}. Moreover, in order to understand whether sympathetic cooling will be sufficient to reach the quantum regime with absorptive particles, the limitations of the technique need to be studied. Finally, sympathetic detection has not previously been demonstrated for mesoscopic particles.

Here we report on sympathetic cooling and detection along all three spatial axes in a system of two levitated mesoscopic particles. We investigate different regimes of the feedback cooling strength and identify the optimum strategy to achieve the minimum temperature of the sympathetically cooled particle. Finally, we reconstruct the motion of a levitated particle via interaction with an auxiliary particle present in the trap. In this proof-of-principle demonstration, both particles are transparent to laser radiation, but it is possible to extend the same techniques to particles with arbitrary optical properties. Optimal detection and cooling along all three axes of motion thus paves the way for experiments with absorptive particles in high and ultra-high vacuum and provides a tool to control the motion of arrays of several particles, which could be used in the search for dark matter~\cite{moore2021searching,carney2021mechanical,afek2022coherent} and in investigations of quantum gravity~\cite{bose2017entanglement}. 

 \begin{figure}[!t]
	\centering
	\includegraphics[width=.9\linewidth]{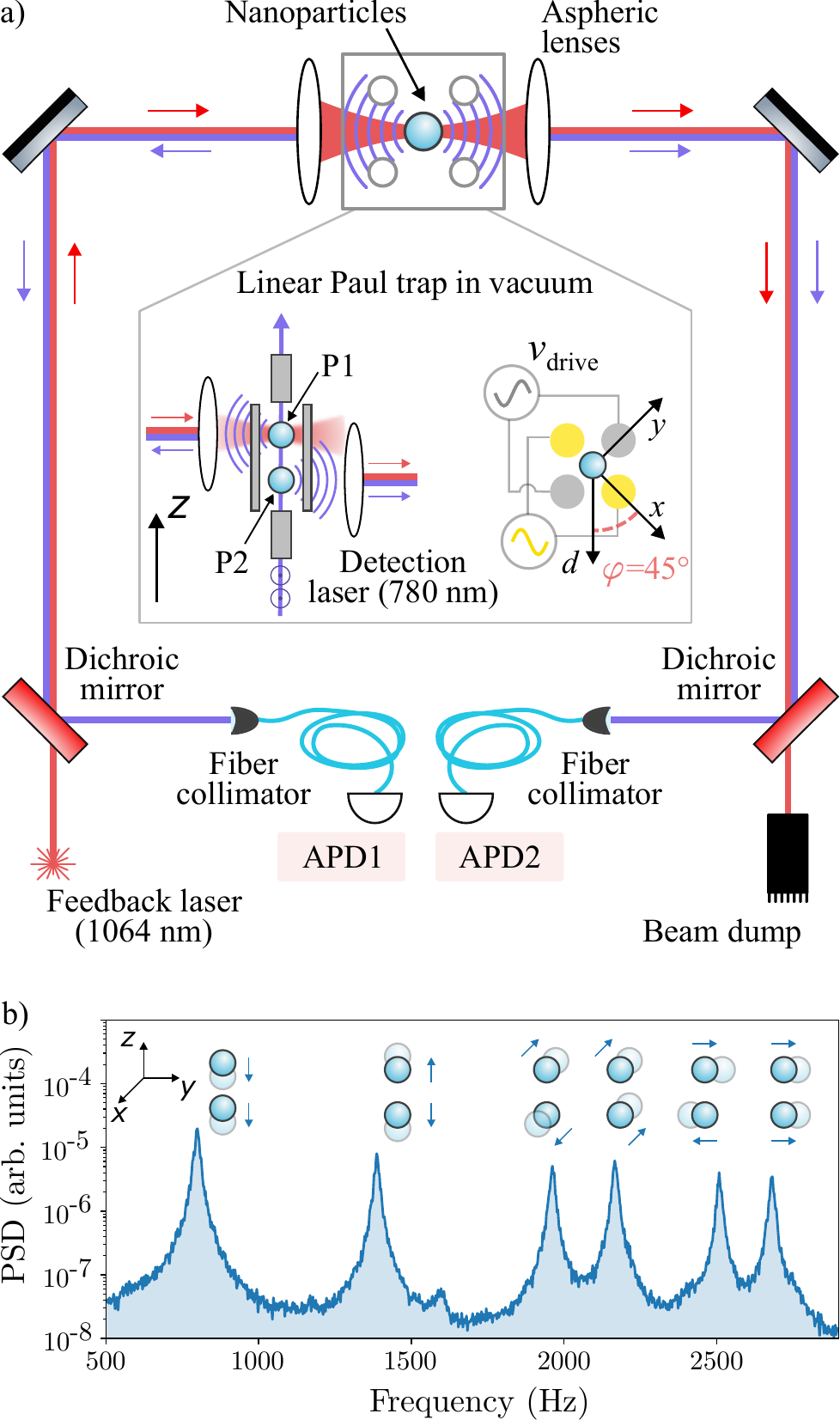}
	\caption{Trapping and detection of two particles. (a) Schematic overview of the experimental setup. The particles are illuminated with a weakly focused laser beam at \SI{780}{\nano\meter} (detection laser). Light from this laser scattered by the particles is collected with aspheric lenses, coupled to optical fibers, and used to detect the individual motion of each nanoparticle at photodiodes APD1 and APD2. A laser beam at \SI{1064}{\nano\meter} (feedback laser) applies an optical feedback force to particle P1. Inset: trap axes and the alignment of the detection system with respect to the particles. (b) An example of the power spectral density (PSD) of the center-of-mass motion of particle P2. Each peak corresponds to one of six collective oscillation modes, illustrated above.}
	\label{fig:fig_1_setup}
\end{figure}
A schematic of the experimental setup is shown in Fig.~\ref{fig:fig_1_setup}a. We levitate a pair of silica spheres \SI{300}{\nano\meter} in diameter (Bangs Laboratories, Inc.) in a linear Paul trap at pressures between \SI{1e-3}{\milli\bar} and \SI{1e-2}{\milli\bar}. Each particle carries \SI{950\pm50} elementary charges~\footnote{See Supplemental Material at [ref. from the journal] for more details on equations of motion, temperature estimation, thermally driven oscillations, particle charging, characterization of the particles, and the transfer function for sympathetic detection, including Refs.~\cite{berkeland98,politzer2015plucked,wubbena2012sympathetic,hebestreit2018calibration,dania2021optical}}. The distance between opposing radiofrequency (RF) electrodes is $2r_0=\SI{3}{\milli\meter}$, while the distance between the endcap electrodes is $2z_0=\SI{7.9}{\milli\meter}$. We drive the trap with a peak-to-peak voltage of \SI{470}{\volt} at \SI{10.1}{\kilo\hertz} and a DC offset on the RF electrodes of \SI{11.5}{\volt}. The DC voltage on the endcaps is varied between \SI{40}{\volt} and \SI{100}{\volt}, depending on the experiment. The endcap voltage sets the distance between the two particles and thus the strength of their electrostatic interaction. 

In order to verify sympathetic cooling and detection, we measure the position of each particle independently via optical detection. The endcap electrodes are hollow cylinders, through which a weakly focused \SI{100}{\milli\watt} laser beam at \SI{780}{\nano\meter} illuminates the particles~(Fig.~\ref{fig:fig_1_setup}a). Light scattered by the particles is collected by a pair of aspheric lenses and guided to fiber collimators. One lens collects light from the top particle, P1, while the second lens collects light from the bottom particle, P2; each lens and collimator image their particle on the end facet of a fiber. The power coupled into each fiber depends on the particle position~\cite{vamivakas2007phasesensitive,xiong2021lensfree} and is measured with avalanche photodiodes (Thorlabs APD440A). A linear relationship between the position of each particle and the photodiode signal is achieved by displacing the image to one side of the fiber core. Cross-talk between the detection channels is below the detection noise.
We choose this detection scheme for its simplicity; we find that it has roughly the same sensitivity (\SI{2e-10}{\meter\per\sqrt{\hertz}}) as that of the standard scheme based on the interference of forward-scattered light and a Gaussian reference beam, although we have not made a detailed comparison of the two schemes' efficiencies.
From the equations of motion of two coupled harmonic oscillators, one finds two eigenmodes of oscillation along each spatial axis $\left\{x,y,z\right\}$, corresponding to the particles' collective in-phase and out-of-phase motion~\cite{Note1}. 

Figure~\ref{fig:fig_1_setup}b shows an example of the power spectral density (PSD) of particle P2's center-of-mass motion, in which each peak corresponds to an eigenmode. The frequencies of these modes are $(\nu_z^{\rm in}, \nu_z^{\rm out},\nu_x^{\rm out}, \nu_x^{\rm in}, \nu_y^{\rm out}, \nu_y^{\rm in}) = (0.8, 1.4, 2.0, 2.2, 2.5, 2.7)\,\si{\kilo\hertz}$. 

A second laser beam at \SI{1064}{\nano\meter} is coupled through a dichroic mirror and focused onto particle P1. The laser power is modulated with an acousto-optic modulator (not shown) controlled via a field-programmable gate array (FPGA). This configuration allows us to apply optical cold damping to P1~\cite{dania2021optical}. 

\begin{figure}[!t]
	\centering
	\includegraphics[width=1\linewidth]{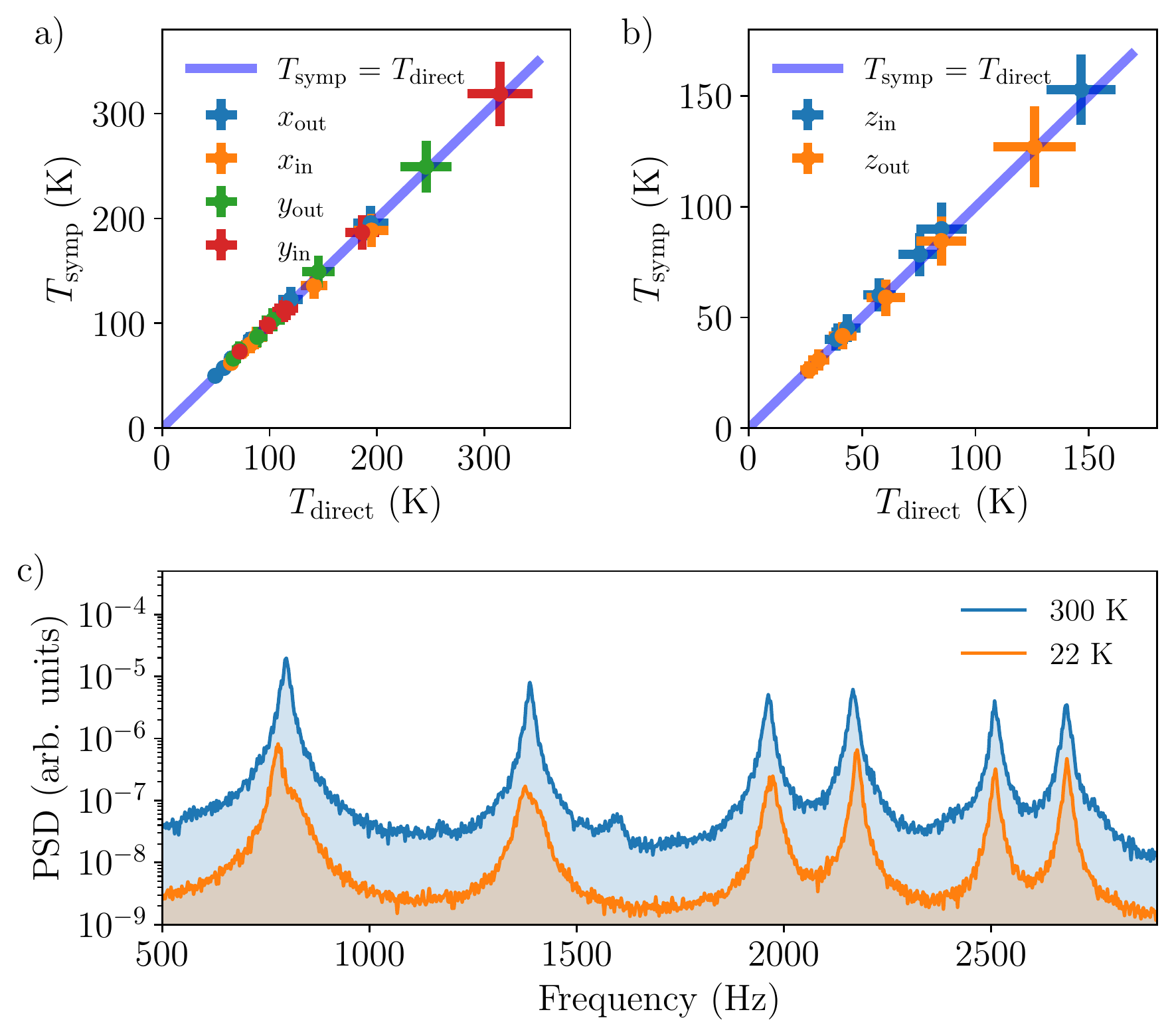}
	\caption{Weak cooling regime. Temperature of the sympathetically cooled particle P2 as a function of the temperature of the directly cooled particle P1 for (a) radial modes $x_{\rm{out}}, x_{\rm{in}}, y_{\rm{out}}, y_{\rm{in}}$ and (b) axial modes $z_{\rm{in}}, z_{\rm{out}}$. The line $T_{\rm{symp}}=T_{\rm{direct}}$ is plotted as a guide to the eye. Error bars correspond to one standard deviation and are determined from the fit. (c) Power spectral densities (PSDs) of the motion of P2 at room temperature (blue curve) and with feedback cooling applied simultaneously to all six collective oscillation modes (orange curve). When feedback cooling is applied, the mean temperature of the six modes is 22 K.
	}
	\label{fig:fig_2_weak}
\end{figure}
Having reviewed the setup, we now present experimental results on sympathetic cooling of particle P2. All data presented in the current manuscript were obtained with the same pair of nanoparticles; the nanoparticles have the same size and the same charge~\cite{Note1}. We start with an analysis of the weak cooling regime. In this case, the rate  $\gamma_{\mathrm{fb}, q}$ of feedback cooling applied to particle P1 along axis $q$ is much smaller than the frequency splitting between the collective oscillation modes: $\gamma_{\mathrm{fb}, q}/2\pi\ll |\nu_q^\mathrm{in}-\nu_q^\mathrm{out}| \approx j_q/2\pi\omega_{q,0}$. Here $q \in \{x,y,z\}$, $j_q$ is a coupling constant and $\omega_{q,0}$ is the resonance frequency of a single particle in the Paul trap along axis $q$~\cite{Note1}. We study the cooling of each mode individually. The motional temperature of P1 is set via the strength of the cold damping, which is controlled by the feedback gain on the FPGA. For each setting of the feedback gain, we determine the temperatures of both P1 (directly cooled) and P2 (sympathetically cooled) by fitting the PSD with a Lorentzian function~\cite{Note1}. Figures~\ref{fig:fig_2_weak}a and \ref{fig:fig_2_weak}b show the dependence of the temperature of P2 on the temperature of P1 for each collective mode. In this regime, we find that the temperatures of the effective CoM motion of the two particles are equal, that is, the temperature of P1 is fully transferred to P2. This result is consistent with the transfer of energy between the particles being significantly faster than the cooling rate of the directly cooled particle. The minimum temperatures shown in Fig.~~\ref{fig:fig_2_weak} are determined by the feedback gain set via the FPGA and by the background gas pressure. We choose to work at \SI{2.7e-3}{\milli\bar} for convenience: this pressure makes the trapping robust against accidental excitation of the motion (and subsequent loss of the particles) while still allowing us to work in the weak cooling regime.
There are no intrinsic limitations to extending sympathetic cooling to lower pressures~\cite{penny2023sympathetic}.

It is possible to sympathetically cool not only individual modes but also all six degrees of freedom simultaneously. To this end, we apply cold damping via six FPGAs, each configured to address a single mode. The six output signals of the FPGAs are mixed and used to modulate the power of the \SI{1064}{\nano\meter} laser. Figure~\ref{fig:fig_2_weak}c compares the PSDs of the sympathetically cooled particle at room temperature and under the influence of feedback cooling. The room-temperature PSD is measured at \SI{2.7e-2}{\milli\bar}, while the PSD of the cooled particle is obtained at \SI{2.3e-3}{\milli\bar}. (We intended to take both data sets at the same pressure, but due to a citywide power failure, the particles were lost from the trap before we could do so.) The mean CoM temperature of the particle under feedback cooling is \SI{22}{\kelvin}; just as for the individual modes, it is determined by the feedback gain and the background gas pressure. 

\begin{figure}[!t]
	\includegraphics[width=1\linewidth]{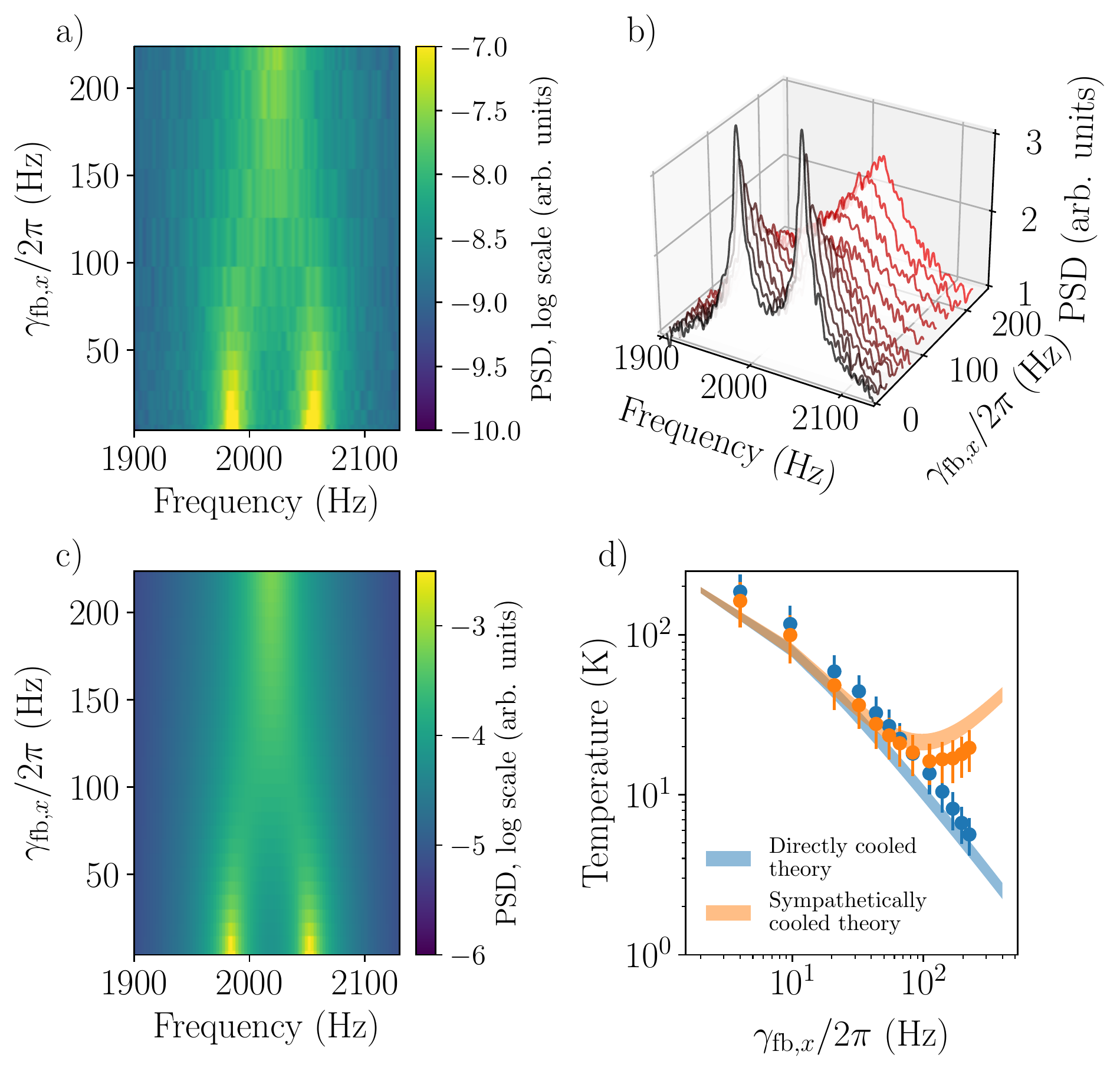}
	\caption{Strong cooling regime. (a)  Power spectral density (PSD) of the motion of particle P2 as a function of the feedback cooling rate, $\gamma_{\mathrm{fb}, x}$. (b) The same data plotted as frequency traces for each value of $\gamma_{\mathrm{fb}, x}$. (c) Calculated PSD of P2 based on a coupled-oscillators model. (d) Temperatures of both particles as a function of the feedback cooling rate. Error bars correspond to one standard deviation and are determined from the uncertainty in the calculation of the area under the velocity PSD curves~\cite{Note1}. The thickness of the theory lines expresses the temperature uncertainty due to the estimation of the gas damping parameter.}
	\label{fig:fig_3_strong_cooling}
\end{figure}
Next, we examine the regime of strong cooling. In this regime, the cooling rate is comparable to or larger than the normal-mode frequency difference: $\gamma_{\mathrm{fb},q}/2\pi \gtrsim |\nu_q^\mathrm{in}-\nu_q^\mathrm{out}| \approx j_q/2\pi\omega_0$. We consider motion along the $x$ axis. By reducing the endcap voltage and thus increasing the particle separation, we reduce the mode splitting to $\nu_x^{\rm in}-\nu_x^{\rm out} \approx \SI{70}{\hertz}$; we then vary the cooling rate from \SI{2}{\hertz} to \SI{220}{Hz} and record PSDs of both particles' motion. The cooling rate is determined as follows: In the weak cooling regime, fits of the mode linewidths allow us to extract cooling rates for a range of electronic gains~\cite{zanette2018energy}. We then have a known relationship between the electronic gain and the cooling rate that also applies in the strong cooling regime.

The PSD of the sympathetically cooled particle P2 as a function of the cooling rate is shown in Figs.~\ref{fig:fig_3_strong_cooling}a and \ref{fig:fig_3_strong_cooling}b. For weak cooling, we observe two distinct peaks of the resonant motion. As the cooling rate is increased into the strong cooling regime, the peaks broaden and merge into a single peak, the height of which increases. A detailed theoretical analysis based on the model of coupled harmonic oscillators~\cite{Note1} shows that the two modes are still present. However, only one of them is not overdamped and thus is still visible in the PSD. We extend the model by coupling the oscillators to a thermal bath~\cite{Note1} and calculate PSDs for different feedback cooling rates. The resulting plot is shown in Fig.~\ref{fig:fig_3_strong_cooling}c. We find good agreement between the measured and the calculated evolution of the PSD, suggesting that the model captures the essential features of the cooling process.

With this understanding of the feedback cooling process in hand, we analyze the particles' temperatures as a function of the cooling rate. Since the two individual modes cannot be resolved in the strong cooling regime, we calculate the effective temperature of motion along the $x$ axis from the area under the velocity PSD~\cite{Note1}. Figure~\ref{fig:fig_3_strong_cooling}d shows the extracted temperatures as a function of the cooling rate applied to the directly cooled particle P1. For cooling rates below the frequency of the normal-mode splitting, the two particles have equal temperatures, similar to the data shown in Fig.~\ref{fig:fig_1_setup}. As the cooling rate increases, the temperature of P1 continues to decrease monotonically, while the temperature of P2 reaches a minimum. We also plot the temperatures extracted from the calculated PSDs of Fig.~\ref{fig:fig_3_strong_cooling}c, which are in agreement with the experimental data.

To gain intuition about the origin of the temperature minimum under sympathetic cooling, one can consider the efficiency of the energy exchange between the two particles as coupled harmonic oscillators. First, this efficiency is limited by the coupling strength. In the strong cooling regime, we extract energy from P1 faster than it flows from P2 to P1. At this point, increasing the cooling rate further does not lower the temperature of P2. Second, two coupled oscillators exchange energy most efficiently when their bare resonance frequencies, i.e., in the absence of coupling, are equal. Damping applied to P1 shifts the particle's bare resonance to a lower frequency than that of P2, making the energy exchange inefficient. As a result, in the strong cooling regime, the flow of energy from P2 to P1 is reduced.

\begin{figure}[!t]
	\includegraphics[width=1\linewidth]{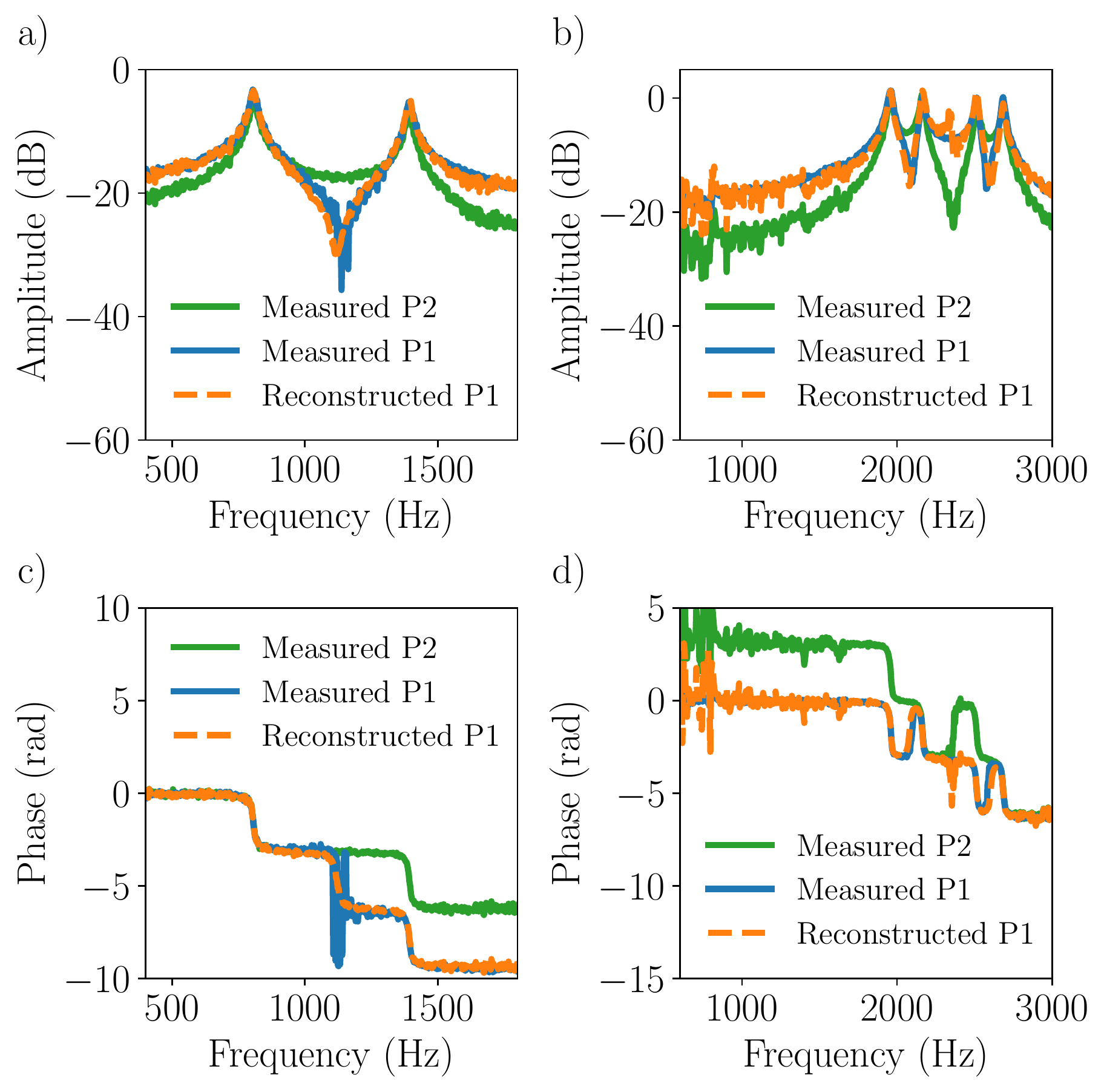}
	\caption{Sympathetic detection. (a) Amplitude response along the $z$ axis. (b) Amplitude response in the $xy$ plane. (c) Phase response along the $z$ axis. (d) Phase response in the $xy$ plane.}
	\label{fig:fig_5_detection}
\end{figure}

Not only can an auxiliary particle in an ion trap be exploited for sympathetic cooling, but also it can be used to detect the motion of the particle of interest, which is especially useful when that motion cannot be read out directly. We implement sympathetic detection, for which the roles of particles P1 and P2 are reversed: P2 is used to detect the motion of P1.
We excite the motion of P1 by applying a sinusoidal optical force to it via the \SI{1064}{\nano\meter} laser beam. No external driving force is applied to P2. We scan the frequency of the force and measure the amplitude and phase response of both particles, which are shown in Fig.~\ref{fig:fig_5_detection}. The amplitude response of P2 to the external drive of P1 allows us to calculate a transfer function $\Theta\left(\omega\right)$ that connects the motion of P1 and P2~\cite{Note1}. From the transfer function, we reconstruct the amplitude and the phase response of P1 from the measurements of P2 via
\begin{equation}
	\begin{split}
		A_\mathrm{P1}\left(\omega\right) = A_\mathrm{P2}\left(\omega\right) / |\Theta\left(\omega\right)| \\
		\phi_\mathrm{P1}\left(\omega\right) = \phi_\mathrm{P2}\left(\omega\right) - \arg \Theta\left(\omega\right),
	\end{split}
\end{equation}
where $\left\{A_\mathrm{P1},A_\mathrm{P2}\right\}$ and $\left\{\phi_\mathrm{P1},\phi_\mathrm{P2}\right\}$ are the amplitude and phase responses of the particles. The reconstructed response functions $A_\mathrm{P1}$ and $\phi_\mathrm{P1}$, plotted in Fig.~\ref{fig:fig_5_detection}, are in excellent agreement with the independently measured values of P1, establishing that sympathetic detection can be used to reconstruct the motion of P1. The dip between the in-phase and out-of-phase modes in the amplitude response of P1 is due to destructive interference of the motion in those modes when one particle is driven, an effect analogous to electromagnetically induced transparency~\cite{alzar2002classical}.

In conclusion, we have demonstrated both sympathetic cooling and detection of nanoparticles levitated in a Paul trap and coupled via electrostatic interaction. If the cooling rate is smaller than the normal-mode frequency difference, the temperature of the sympathetically cooled particles equals the temperature of the directly cooled particle. For higher cooling rates, the temperature of the sympathetically cooled particle reaches a minimum value due to reduced efficiency of the energy exchange between two coupled harmonic oscillators. Both methods pave the way for experiments with absorptive particles in high and ultra-high vacuum in the absence of light. For example, a nanodiamond with nitrogen vacancies could be co-trapped with a silica nanoparticle; both particles' charge and mass could be calibrated at low vacuum, where the diamond is not significantly heated by laser light; subsequently, at ultra-high vacuum, the silica nanoparticle could be used to measure and control the motion of the nanodiamond. Furthermore, the technique provides a tool to detect and control the $3N$ modes of collective oscillations of a string of $N$ particles by acting on just a single particle in the chain.
Building on the spin-mechanical coupling recently demonstrated with a single levitated diamond hosting nitrogen-vacancy defects~\cite{delord2020spincooling}, our work opens up new experimental possibilities for realizing a quantum spin transducer based on levitated particles. Such particles could be used as long-lived quantum memories for quantum information applications~\cite{Rabl_2010,Rosenfeld_2021} or for force sensing below the standard quantum limit in the search for new physics~\cite{moore2021searching,carney2021mechanical,afek2022coherent}.	

\begin{acknowledgments}
This work was supported by Austrian Science Fund (FWF) Projects No. W1259 and Y951, by the ESQ Discovery Grant “Sympathetic detection and cooling of nanoparticles levitated in a Paul trap” of the Austrian Academy of Sciences, and by the European Union’s Horizon 2020 research
and innovation program under the Marie Skłodowska-Curie Grant Agreement No. 801110.

Data underlying the results of this study are available at~\cite{bykov2022sympathetic_data}.
\end{acknowledgments}

\bibliography{biblio_sympathetic}
\end{document}


\title{Supplemental material for ``3D sympathetic cooling and detection of levitated nanoparticles"}
\author{Dmitry S. Bykov}
\email[]{dmitry.bykov@uibk.ac.at}
\affiliation{Institut f{\"u}r Experimentalphysik, Universit{\"a}t Innsbruck, Technikerstra\ss e 25, 6020 Innsbruck,
	Austria}
\author{Lorenzo Dania}
\affiliation{Institut f{\"u}r Experimentalphysik, Universit{\"a}t Innsbruck, Technikerstra\ss e 25, 6020 Innsbruck,
	Austria}
\author{Florian Goschin}
\affiliation{Institut f{\"u}r Experimentalphysik, Universit{\"a}t Innsbruck, Technikerstra\ss e 25, 6020 Innsbruck,
	Austria}
\author{Tracy E. Northup}
\affiliation{Institut f{\"u}r Experimentalphysik, Universit{\"a}t Innsbruck, Technikerstra\ss e 25, 6020 Innsbruck,
	Austria}

\date{\today}
\maketitle
\onecolumngrid

\appendix
\section{Equations of motion}
\label{app:eoms}
\begin{figure}[t]
	\centering
	\includegraphics[width=0.3\linewidth]{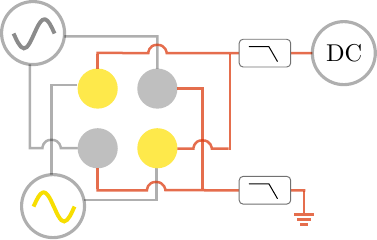}
	\caption{Schematic of the voltages applied to the RF electrodes. An AC voltage is applied to all four electrodes with the same frequency and amplitude; the signal applied to two electrodes is shifted in phase by \ang{180}. A DC offset is applied to two of the four electrodes.}
	\label{app_fig:drive}
\end{figure}

In this section, the motion of nanoparticles is analyzed in two cases: free evolution and driven oscillations. We model two levitated and electrostatically interacting nanoparticles as coupled harmonic oscillators. We assume that feedback cooling is applied to the first particle, referred to in the main text as P1 and here denoted with 1 in subscript. For displacements that are small compared to the distance between the particles, the equations of motion can be written as
\begin{equation}\label{eq:motion_matrix_form}
	\frac{d}{dt}
	\begin{bmatrix}
		{q}_{1} \\
		{q}_{2} \\
		\dot{q}_{1} \\
		\dot{q}_{2}
	\end{bmatrix}
	=
	\boldsymbol{M}
	\begin{bmatrix}
		q_{1} \\
		q_{2} \\
		\dot{q}_{1} \\
		\dot{q}_{2}
	\end{bmatrix}
	+
	\begin{bmatrix}
		0 \\
		0 \\
		F_{q,1}/m_1 \\
		F_{q,2}/m_2
	\end{bmatrix}
\end{equation}
with
\begin{equation}\label{eq:matrix}
	\boldsymbol{M}  =
	\begin{bmatrix}
		0 & 0 & 1 & 0\\
		0 & 0 & 0 & 1\\
		-\omega_{0q,1}^2 + j_{q} & -j_{q} & -\gamma_{01} - \gamma_{\mathrm{fb},q} & 0\\
		-\mu j_{q} & -\omega_{0q,2}^2 + \mu j_{q} & 0 & -\gamma_{02}
	\end{bmatrix},
\end{equation}
where $q \in \{x,y,z\}$ is the axis of motion, and the particles at rest lie along the $z$ axis; $m_{1}$ and $m_{2}$ are the masses of the nanoparticles; $F_{q,1}$ and $F_{q,2}$ are the external forces acting on the first and second particle, respectively;  $\omega_{0q,1}$ and $\omega_{0q,2}$ are the resonance frequencies of each particle's center-of-mass (CoM) motion; $\gamma_{01}$ and  $\gamma_{02}$ are the rates at which the particles are damped by background gas, which depend on their radii and mass; $\gamma_{\mathrm{fb},q}$ is the cooling rate; $\mu = m_1/m_2$ is the mass ratio; and $j_{q}$ is a coupling constant. For axes $x$ and $y$, the coupling constant is $j_{x} = j_{y} = Q_1 Q_2 / \left(4\pi\epsilon_0 d^3 m_{1}\right)$; for axis $z$, the coupling constant is $j_{z}=-2Q_1 Q_2 / 4\pi\epsilon_0 d^3 m_{1}$, where $Q_{1}$ and $Q_{2}$ are the particles' charges and $d$ is the distance between the particles. The resonance frequencies $\omega_{0q}$ of a charged particle in a linear Paul trap can be calculated as~\cite{berkeland98}
\begin{align}
	&\omega_{0x} \cong \frac{\Omega_\text{d}}{2} \sqrt{a_x + q_x^2/2}, \label{eq:omegax}\\
	&\omega_{0y} \cong \frac{\Omega_\text{d}}{2} \sqrt{a_y + q_y^2/2}, \label{eq:omegay}\\
	&\omega_{0z} \cong \frac{\Omega_\text{d}}{2}\sqrt{a_z}, \label{eq:omegaz}
\end{align}
with the $a$ and $q$ parameters defined as
\begin{align}
	&a_x = -\frac{4Q}{m\Omega_\text{d}^2}\left(\kappa_{\text{end}} \frac{V_{\text{end}}}{z_0^2} + \kappa_{\text{RF}} \frac{V_{\text{off}}}{r_0^2} \right), \\
	&a_y = -\frac{4Q}{m\Omega_\text{d}^2}\left(\kappa_{\text{end}} \frac{V_{\text{end}}}{z_0^2} - \kappa_{\text{RF}} \frac{V_{\text{off}}}{r_0^2} \right), \\
	&a_z = \kappa_{\text{end}} \frac{8Q V_{\text{end}}}{m z_0^2 \Omega_\text{d}^2}, \label{eq:az}\\
	&q_x = \kappa_{\text{RF}} \frac{4QV_{\text{pp}}}{mr_0^2\Omega_\text{d}^2}, \label{eq:qx} \\
	&q_y = -q_x,
\end{align}
where $\Omega_d$ is the frequency of the AC voltage applied to the trap radiofrequency (RF) electrodes, $V_{\text{pp}}$ is the peak-to-peak amplitude of the AC voltage, $V_{\text{off}}$ is the DC offset applied to the RF electrodes, $V_{\text{end}}$ is the DC voltage applied to the trap endcap electrodes, $\kappa_{\text{RF}}$ and $\kappa_{\text{end}}$ are geometrical factors,  $r_0$ is half the distance between opposing RF electrodes, and $z_0$ is half the distance between the endcap electrodes. Our parameters are $V_{\text{pp}} =\SI{470}{\volt}$, $V_{\text{off}} = \SI{11.6}{\volt}$, $\kappa_{\text{RF}} = 0.77$, $\kappa_{\text{end}}=0.25$, $z_0 = \SI{4}{\milli\meter}$, $r_0 = \SI{1.5}{\milli\meter}$, $m = \SI{2e-17}{\kilogram}$, and $Q = 950e$. In our experimental setup, the AC voltage is applied to all four RF electrodes but with a phase shift of $\pi$ between pairs of electrodes, often referred to as an RF-RF configuration, as depicted in Fig.~\ref{app_fig:drive}.
The DC offset $V_{\text{off}}$ is applied to one pair of electrodes.

\FloatBarrier

\subsection{Free oscillations}\label{app:sec_modes}
In the absence of external forces, that is, for $F_{q,1} = F_{q,2} = 0$, the collective motion consists of damped oscillations $\left[ q_1, q_2, \dot{q_1},\dot{q_2}\right]^T = \vec{e}\exp\left(\lambda_q t\right)$ with four complex eigenfrequencies $\lambda_q \in \left\{\lambda^{\pm}_{q,\text{in}}, \lambda^{\pm}_{q,\text{out}}\right\} $, defined as the eigenvalues of matrix $M$ in Eq.~\ref{eq:matrix}, and corresponding eigenvectors $\vec{e} \in \left\{ \left[ a_{q,\mathrm{in}}, b_{q,\mathrm{in}}, \lambda^{\pm}_{q,\text{in}}a_{q,\mathrm{in}}, \lambda^{\pm}_{q,\text{in}}b_{q,\mathrm{in}}\right]^T, \left[ a_{q,\mathrm{out}}, b_{q,\mathrm{out}}, \lambda^{\pm}_{q,\text{out}}a_{q,\mathrm{out}}, \lambda^{\pm}_{q,\text{out}}b_{q,\mathrm{out}}\right]^T \right\} $. The index ``in" corresponds to the mode of collective oscillations with both oscillators moving in phase, while the index ``out" corresponds to the out-of-phase mode. We can write the complex eigenfrequencies in terms of their real and imaginary parts:
\begin{align}
	\lambda^{\pm}_{q,\text{in}} &= -\gamma_{q,\text{in}} \pm i \omega_{q,\text{in}}, \\
	\lambda^{\pm}_{q,\text{out}} &= -\gamma_{q,\text{out}} \pm i \omega_{q,\text{out}}.
\end{align}
The real parts set the damping rate, and the imaginary parts determine the frequency of oscillation. Without loss of generality, we consider only eigenvalues with a positive imaginary part. Our system thus has six modes of collective oscillation, two along each spatial axis.

\begin{figure}[t]
	\centering
	\includegraphics[width=1\linewidth]{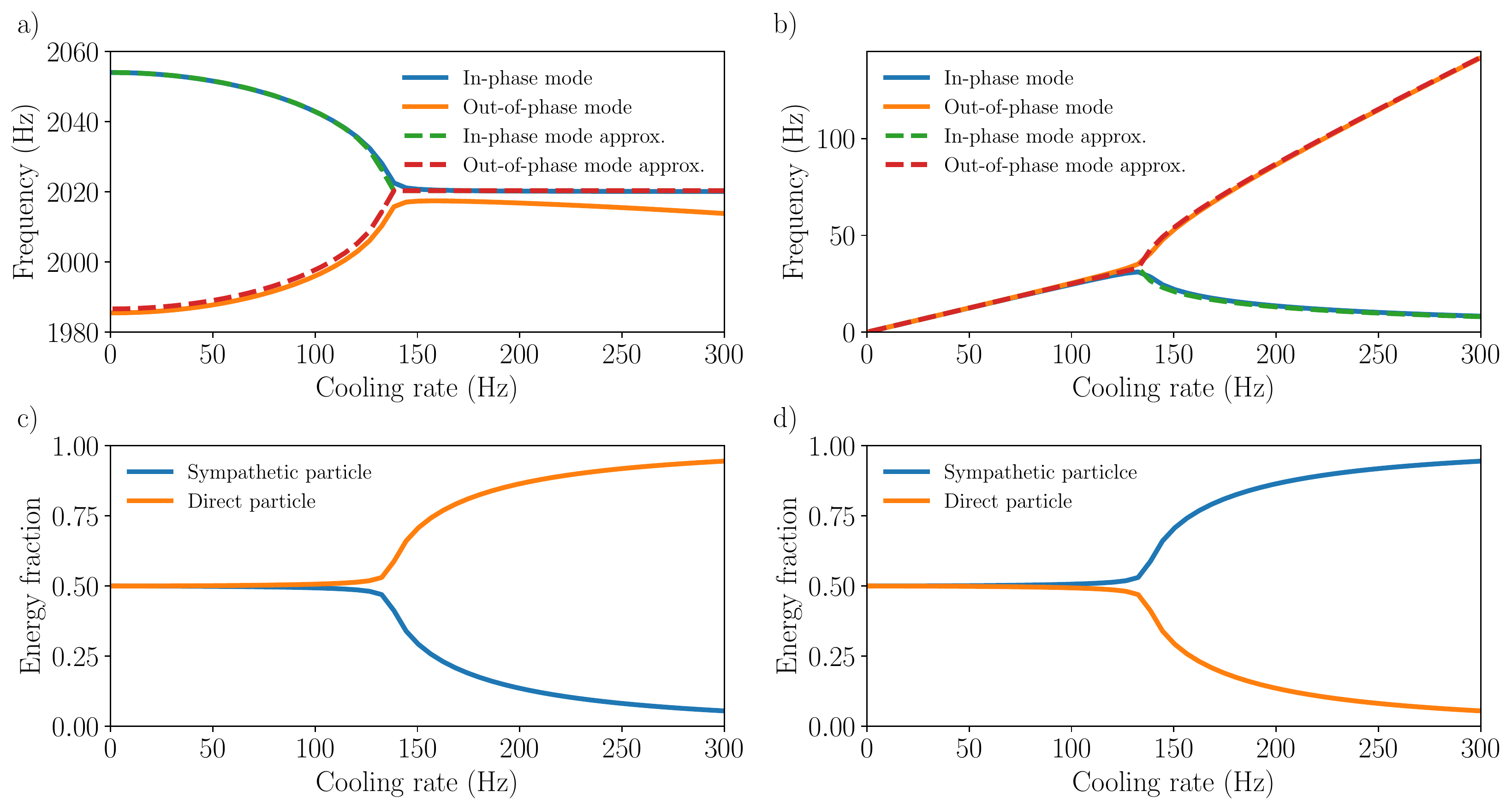}
	\caption{Eigenvalues and eigenvectors of two coupled harmonic oscillators. (a) Imaginary parts of the complex eigenfrequencies as a function of the feedback cooling rate. (b) Real parts of the complex eigenfrequencies as a function of the feedback cooling rate. (c) Normalized energy fractions of the sympathetically cooled particle and the directly cooled particle in the out-of-phase mode. (d) Normalized energy fractions of the sympathetically cooled particle and the directly cooled particle in the in-phase mode.}
	\label{app_fig:fig_1_modes}
\end{figure}
The eigenvalues of matrix $M$ are solutions of a quartic algebraic equation and have cumbersome expressions. We calculate them with a script in Wolfram Mathematica (version 12.1). For identical particles, meaning $\omega_{0x,1} = \omega_{0x,2} = \omega_{0x}$,  $\gamma_{01} = \gamma_{02} = \gamma_{\rm gas}$ and $\mu = 1$, and under the conditions $\gamma_{\mathrm{fb},x} + \gamma_{\mathrm{gas}}\ll\omega_{0x}$ and $j/2\omega_{0x}\ll \omega_{0x}$, the frequencies can be approximated by~\cite{politzer2015plucked}
	\begin{equation}
		\label{app_eq:eigen_values_approx}
		\begin{split}
			\lambda_{x,\mathrm{\mathrm{in}}} &= i\left(\omega_{0x} - \frac{j}{2\omega_{0x}} - i\frac{\gamma_{\mathrm{fb},x} + \gamma_{\mathrm{gas}}}{4}+\sqrt{\left(\frac{j}{2\omega_{0x}}\right)^2-\left(\frac{\gamma_{\mathrm{fb},x} + \gamma_{\mathrm{gas}}}{4}\right)^2}\right)\\
			\lambda_{x,\mathrm{\mathrm{out}}} &= i\left(\omega_{0x} - \frac{j}{2\omega_{0x}} - i\frac{\gamma_{\mathrm{fb},x} + \gamma_{\mathrm{gas}}}{4} - \sqrt{\left(\frac{j}{2\omega_{0x}}\right)^2-\left(\frac{\gamma_{\mathrm{fb},x} + \gamma_{\mathrm{gas}}}{4}\right)^2}\right)
		\end{split}
	\end{equation}
Instead of analyzing the general expression, we focus on the modal structure of the system of two nanoparticles used in the experiments described in the main text. For comparison with the data presented in Fig.~3 of the main text, we restrict ourselves to motion along the $x$ axis. The system parameters are given in Table~\ref{app_table:experimental_parameters}.
\begin{table}[H]
	\begin{center}
		\begin{tabular}{|c|c|}
			\hline
			Parameter & Value\\
			\hline
			\hline
			$\omega_{0x,1} = \omega_{0x}$ & $2\pi\times\SI{2054}{\hertz}$\\
			\hline
			$\omega_{0x,2} = \omega_{0x}$ & $2\pi\times\SI{2054}{\hertz}$\\
			\hline
			$\gamma_{01}=\gamma_{\rm gas}$ & $2\pi\times\SI{1.6}{\hertz}$\\
			\hline
			$\gamma_{02} =\gamma_{\rm gas}$ & $2\pi\times\SI{1.6}{\hertz}$\\
			\hline
			$\sqrt{j}$ & $2\pi\times\SI{372}{\hertz}$\\
			\hline
			$m_1 = m_2$ & \SI{2e-17}{\kilo\gram}\\
			\hline
			$\mu$ & 1\\
			\hline
		\end{tabular}
	\end{center}
\caption{Parameters that characterize the experiment in the main text and that are used in the theoretical model of Appendix~\ref{app:eoms}.}\label{app_table:experimental_parameters}
\end{table} 
The evolution of the modal frequencies $\omega_{x,\text{in}} / 2\pi$ and $\omega_{x,\text{out}} / 2\pi$ is plotted in Fig.~\ref{app_fig:fig_1_modes}a for cooling rates up to $\gamma_{\mathrm{fb},x} = \SI{300}{\hertz}$. We plot both the approximate solutions \ref{app_eq:eigen_values_approx} and the exact eigenvalues of $M$ in Eq.~\ref{eq:matrix}. As $\gamma_{\mathrm{fb},x}$ is increased, the frequency of the out-of-phase mode grows, while the frequency of the in-phase mode drops. When $\gamma_{\mathrm{fb},x}$ reaches the value $2 j_x / \omega_{0x}=\SI{135}{\hertz}$, the approximate solutions yield equal frequencies for the in-phase and out-of-phase modes. The exact solutions show that the frequency of the in-phase mode stabilizes at $\omega_{0x} - j_x/2 \omega_{0x} = 2\pi\times\SI{2020}{\hertz}$, while the frequency of the out-of-phase mode approaches this value but then starts to decrease and reaches \SI{0}{\hertz} in the limit of an infinite feedback cooling rate. The evolution of the modal damping rates $\gamma_{x,\mathrm{in}}$ and $\gamma_{x,\mathrm{out}}$ is shown in Fig.~\ref{app_fig:fig_1_modes}b. For values of $\gamma_{\mathrm{fb},x}$ below $2 j_x / \omega_{0x}$, the modal damping rates are equal and increase monotonically. For values of $\gamma_{\mathrm{fb},x}$ above $2 j_x / \omega_{0x}$, the damping rate of the out-of-phase mode continues to increase while the damping rate of the in-phase mode approaches $\gamma_{\text{gas}}$.

Next, we analyze the energies of the individual particles. Each particle contains some fraction of the energies of the in-phase and out-of-phase modes. These fractions can be calculated as~\cite{wubbena2012sympathetic}
\begin{align}
		D_{\text{in}} &= \frac{a^{2}_{x,\text{in}}}{a^{2}_{x,\text{in}} + b^{2}_{x,\text{in}}},\\
		D_{\text{out}} &= \frac{a^{2}_{x,\text{out}}}{a^{2}_{x,\text{out}} + b^{2}_{x,\text{out}}},\\
		I_{\text{in}} &= \frac{b^{2}_{x,\text{in}}}{a^{2}_{x,\text{in}} + b^{2}_{x,\text{in}}}, \\
		I_{\text{out}} &=  \frac{b^{2}_{x,\text{out}}}{a^{2}_{x,\text{out}} + b^{2}_{x,\text{out}}},
\end{align}
where $D$ is the energy fraction in the directly cooled particle; $I$ is the energy fraction in the sympathetically cooled particle. Figures~\ref{app_fig:fig_1_modes}c and \ref{app_fig:fig_1_modes}d show the energy fractions in the out-of-phase and in-phase modes. As the cooling rate is increased, the energy of the directly cooled particle becomes concentrated in the out-of-phase mode, while the energy of the sympathetically cooled particle becomes concentrated in the in-phase mode.

\subsection{Driven oscillations}
In the presence of an external drive, that is, for $F_{q,1} \neq 0$ or $F_{q,2} \neq 0$, the steady-state solution of Eq.~\ref{eq:motion_matrix_form} in the frequency domain can be written as
\begin{equation}\label{eq:motion_steady}
	\begin{bmatrix}
		q_{1}\left(\omega\right) \\
		q_{2}\left(\omega\right)
	\end{bmatrix}
	=
	\begin{bmatrix}
		\chi_{q,11}\left(\omega\right) & \chi_{q,12}\left(\omega\right)\\
		\chi_{q,21}\left(\omega\right) & \chi_{q,22}\left(\omega\right)
	\end{bmatrix}
	\begin{bmatrix}
		F_{q,1}\left(\omega\right) \\
		F_{q,2}\left(\omega\right)
	\end{bmatrix}
\end{equation}
with response functions $\chi(\omega)$ defined as
\begin{align}
	\label{eq:transfer_x11}
	\chi_{x,11}\left(\omega\right) &=  \frac{1}{m_{1}}\frac{j_{x,2} + i \omega \gamma'_{02} + \omega^2 - \omega_{0x,2}^2}{j_{x,1}j_{x,2} - \left(j_{x,1} + i\omega \gamma'_{01} + \omega^2 -\omega_{0x,1}^2\right)\left(j_{x,2} + i\omega \gamma'_{02} + \omega^2 -\omega_{0x,2}^2\right)}\\
	\label{eq:transfer_x22}
	\chi_{x,22}\left(\omega\right) &=  \frac{1}{m_{2}}\frac{j_{x,1} + i \omega \gamma'_{01} + \omega^2 - \omega_{0x,1}^2}{j_{x,1}j_{x,2} - \left(j_{x,1} + i\omega \gamma'_{01} + \omega^2 -\omega_{0x,1}^2\right)\left(j_{x,2} + i\omega \gamma'_{02} + \omega^2 -\omega_{0x,2}^2\right)}\\
	\label{eq:transfer_y11}
	\chi_{y,11}\left(\omega\right) &=  \frac{1}{m_{1}}\frac{j_{y,2} + i \omega \gamma'_{02} + \omega^2 - \omega_{0y,2}^2}{j_{y,1}j_{y,2} - \left(j_{y,1} + i\omega \gamma'_{01} + \omega^2 -\omega_{0y,1}^2\right)\left(j_{y,2} + i\omega \gamma'_{02} + \omega^2 -\omega_{0y,2}^2\right)}\\
	\label{eq:transfer_y22}
	\chi_{y,22}\left(\omega\right) &=  \frac{1}{m_{2}}\frac{j_{y,1} + i \omega \gamma'_{01} + \omega^2 - \omega_{0y,1}^2}{j_{y,1}j_{y,2} - \left(j_{y,1} + i\omega \gamma'_{01} + \omega^2 -\omega_{0y,1}^2\right)\left(j_{y,2} + i\omega \gamma'_{02} + \omega^2 -\omega_{0y,2}^2\right)}\\
	\label{eq:transfer_x21}
	\chi_{x,21}\left(\omega\right) &= \frac{1}{m_{1}}\frac{j_{x,2}}{j_{x,1}j_{x,2} - \left(j_{x,1} + i\omega \gamma'_{01} + \omega^2 -\omega_{0x,1}^2\right)\left(j_{x,2} + i\omega \gamma'_{02} + \omega^2 -\omega_{0x,2}^2\right)}\\
	\label{eq:transfer_x12}
	\chi_{x,12}\left(\omega\right) &= \frac{1}{m_{2}}\frac{j_{x,1}}{j_{x,1}j_{x,2} - \left(j_{x,1} + i\omega \gamma'_{01} + \omega^2 -\omega_{0x,1}^2\right)\left(j_{x,2} + i\omega \gamma'_{02} + \omega^2 -\omega_{0x,2}^2\right)}\\
	\label{eq:transfer_y21}
	\chi_{y,21}\left(\omega\right) &= \frac{1}{m_{1}}\frac{j_{y,2}}{j_{y,1}j_{y,2} - \left(j_{y,1} + i\omega \gamma'_{01} + \omega^2 -\omega_{0y,1}^2\right)\left(j_{y,2} + i\omega \gamma'_{02} + \omega^2 -\omega_{0y,2}^2\right)}\\
	\label{eq:transfer_y12}
	\chi_{y,12}\left(\omega\right) &= \frac{1}{m_{2}}\frac{j_{y,1}}{j_{y,1}j_{y,2} - \left(j_{y,1} + i\omega \gamma'_{01} + \omega^2 -\omega_{0y,1}^2\right)\left(j_{y,2} + i\omega \gamma'_{02} + \omega^2 -\omega_{0y,2}^2\right)}\\
	\label{eq:transfer_z11}
	\chi_{z,11}\left(\omega\right) &= \frac{1}{m_{1}} \frac{-j_{z,2} + i \omega \gamma'_{02)} + \omega^2 - \omega_{0z,2}^2}{j_{z,1}j_{z,2} - \left(j_{z,1} - i\omega \gamma'_{01} - \omega^2 +\omega_{0z,1}^2\right)\left(j_{z,2} - i\omega \gamma'_{02} - \omega^2 +\omega_{0z,2}^2\right)}\\
	\label{eq:transfer_z22}
	\chi_{z,22}\left(\omega\right) &= \frac{1}{m_{2}} \frac{-j_{z,2} + i \omega \gamma'_{01} + \omega^2 - \omega_{0z,1}^2}{j_{z,1}j_{z,2} - \left(j_{z,1} - i\omega \gamma'_{01} - \omega^2 +\omega_{0z,1}^2\right)\left(j_{z,2} - i\omega \gamma'_{02} - \omega^2 +\omega_{0z,2}^2\right)}\\
	\label{eq:transfer_z21}
	\chi_{z,21}\left(\omega\right) &= \frac{1}{m_{1}} \frac{-j_{z,2}}{j_{z,1}j_{z,2} - \left(j_{z,1} - i\omega \gamma'_{01} - \omega^2 + \omega_{0z1}^2\right)\left(j_{z,2} - i\omega \gamma'_{02} - \omega^2 +\omega_{0z,2}^2\right)}\\
	\label{eq:transfer_z12}
	\chi_{z,12}\left(\omega\right) &= \frac{1}{m_{2}} \frac{-j_{z,1}}{j_{z,1}j_{z,2} - \left(j_{z,1} - i\omega \gamma'_{01} - \omega^2 + \omega_{0z1}^2\right)\left(j_{z,2} - i\omega \gamma'_{02} - \omega^2 +\omega_{0z,2}^2\right)},
\end{align}
for $j_{q,2} = \mu j_{q,1}$, $\gamma'_{01}=\gamma_{01} + \gamma_{\mathrm{fb},q}$  and $\gamma'_{02}=\gamma_{02}$.

\section{Temperature estimation}
This section describes how to extract the temperature of the CoM motion of the particles along axis $q$ using two different methods~\cite{hebestreit2018calibration}. The first method is based on fitting power spectral densities (PSDs) $ S_{qq,1}\left(\omega\right)$ and $S_{qq,2}\left(\omega\right)$ with Lorentzian functions, while the second method relies on calculating the area under the curves $S_{\dot{q}\dot{q},1}\left(\omega\right) = \omega^2 S_{qq,1}\left(\omega\right)$ and $S_{\dot{q}\dot{q},2}\left(\omega\right) = \omega^2 S_{qq,2}\left(\omega\right)$, where the subscripts 1 and 2 indicate the particle number. We use the first method to calculate the temperatures of the individual modes and the second method to obtain the total kinetic energy of a single particle along a given spatial axis $x$, $y$, or $z$. In the latter case, the motion comprises two modes: the in-phase mode and the out-of-phase mode.

\subsection{Temperature estimation of a mode of collective oscillation}
In the weak cooling regime, defined as $\gamma_{\mathrm{fb},q}\ll j_q/ \omega_{0q,1}$ and $\gamma_{\mathrm{fb},q}\ll j_q/ \omega_{0q,2}$, we can consider each mode as an independent oscillator. Each PSD of particle motion can be written as the sum of two Lorentzian functions:
\begin{align}
	\label{app_eq:PSD1}
	S_{qq,1}\left(\omega\right) &= \left(\overbrace{S^{\text{(th)}}_{q,\mathrm{in}} \frac{1}{4m_\mathrm{eff,in}^2}\frac{1}{\gamma_\mathrm{0,in}^2\omega^2 + \left(\omega_{q,\mathrm{in}}^2-\omega^2\right)^2}}^{\text{in-phase mode}} + \overbrace{S^{\mathrm{(th)}}_{q,\mathrm{out}} \frac{1}{4m_\mathrm{eff,out}^2}\frac{1}{1+\mu^2}\frac{1}{\gamma_\mathrm{0,out}^2\omega^2 + \left(\omega_{q,\mathrm{out}}^2-\omega^2\right)^2}}^{\text{out-of-phase mode}}\right),\\
	\label{app_eq:PSD2}
	S_{qq,2}\left(\omega\right) &= \left(S^{\text{(th)}}_{q,\mathrm{in}} \frac{1}{4m_\mathrm{eff,in}^2}\frac{1}{\gamma_\mathrm{0,in}^2\omega^2 + \left(\omega_{q,\mathrm{in}}^2-\omega^2\right)^2} + S^{\text{(th)}}_{q,\mathrm{out}} \frac{1}{4m_\mathrm{eff,out}^2}\frac{\mu^2}{1+\mu^2}\frac{1}{\gamma_\mathrm{0,out}^2\omega^2 + \left(\omega_{q,\mathrm{out}}^2-\omega^2\right)^2}\right),
\end{align}
where
\begin{align}
	S^{\text{(th)}}_{q,\mathrm{in}} &= 4 \left(\frac{\gamma_{01} + \gamma_{02}}{2}\right) \frac{m_\mathrm{eff,in} k_{\text{B}} T}{\pi}\\
	S^{\text{(th)}}_{q,\mathrm{out}} &= 4 \left(\frac{\gamma_{01} + \gamma_{02}}{2}\right) \frac{m_\mathrm{eff,out} k_{\text{B}} T}{\pi}
\end{align}
are the PSDs of the thermal force acting on each mode, and
\begin{align}
	m_\mathrm{eff,in} &= \frac{m_{1} + m_{2}}{2}\\
	m_\mathrm{eff,out} &= \frac{m_{1}m_{2}}{m_{1} + m_{2}}
\end{align}
are the effective masses of the modes. We extract the modal temperature of each particle by fitting the PSD for each mode to a Lorentzian~\cite{hebestreit2018calibration,dania2021optical}. While the measured PSDs can be expressed in units of \SI{}{\meter^2\hertz^{-1}}, it is not necessary for temperature estimation, so we use arbitrary units for power spectral density plots.

\subsection{Temperature estimation of the center of mass of an individual particle}
If the individual modes of collective oscillations cannot be resolved, for example, when the damping of the motion is larger than the normal mode splitting, the particle's temperature along axis $q$ can be calculated based on the average kinetic energy~\cite{hebestreit2018calibration}
\begin{equation}\label{eq:kinetic}
	\langle E_\mathrm{kin} \rangle = \frac{1}{2}m \langle \dot{q}^2 \rangle = \frac{1}{2}k_B T,
\end{equation}
The variance $\langle \dot{q}^2 \rangle$ can be calculated as $\langle \dot{q}^2 \rangle = \int_{0}^{\infty} S_{\dot{q}\dot{q}}(\omega) \,d\omega  = \int_{0}^{\infty}\omega^2 S_{qq}(\omega) \,dx$, where $S_{\dot{q}\dot{q}}(\omega)$ is the velocity PSD and $S_{qq}(\omega)$ is the displacement PSD. However, since the experimentally determined $S_{qq}$ contains a detection-noise component, integration over all frequencies will lead to an overestimate of the velocity variance. Therefore, we limit the integration range to frequencies at which the detection signal is dominated by the particle motion.

\section{Thermally driven oscillations}
\begin{figure}[H]
	\centering
	\includegraphics[width=0.8\linewidth]{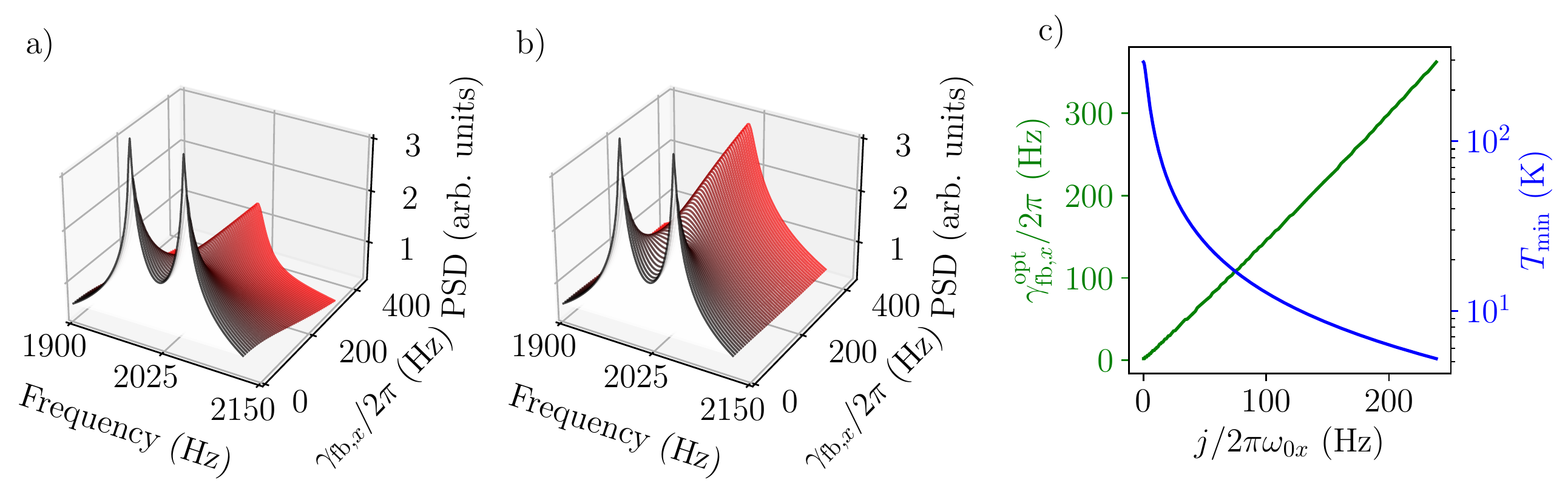}
	\caption{Thermally driven oscillations of cooled particles. Theoretical PSDs of (a) the directly cooled particle and (b) the sympathetically cooled particle, as a function of the feedback cooling rate $\gamma_{{\rm{fb}},x}$. (c) Optimum feedback cooling rate $\gamma_{{\rm{fb}},x}^{\rm{opt}}$ and minimum temperature of the sympathetically cooled particles $T_{\rm min}$as a function of the normal-mode splitting parameter $j/2\pi\omega_{0x}$.}
	\label{app_fig:fig_2_theory_psd}
\end{figure}
This section studies the case of the external drive provided by a random thermal force. We analyse PSDs and temperature of both particles as a function of the feedback cooling rate $\gamma_{\mathrm{fb},x}$ along axis $x$.
As in Section~\ref{app:sec_modes}, the parameters we use in the calculation --- listed in Table~\ref{app_table:experimental_parameters} --- characterize the experiment for which data are shown in Fig.~3 of the main text. We continue to denote the directly cooled particle with index 1 and the sympathetically cooled particle with index 2. The rate of feedback cooling $\gamma_{\mathrm{fb},x}$ applied to the first particle is a variable parameter. 

Under the influence of a random thermal force, the PSDs $S_{qq,1}$ and $S_{qq,2}$ of the motion of the particles can be written as
\begin{equation}\label{eq:psd}
	\begin{bmatrix}
		S_{qq,1}\left(\omega\right) \\
		S_{qq,2}\left(\omega\right)
	\end{bmatrix}
	=
	\begin{bmatrix}
		\left|\chi_{q,11}\left(\omega\right)\right|^2 & \left|\chi_{q,12}\left(\omega\right)\right|^2\\
		\left|\chi_{q,21}\left(\omega\right)\right|^2 & \left|\chi_{q,22}\left(\omega\right)\right|^2
	\end{bmatrix}
	\begin{bmatrix}
		S^{\text{(th)}}_{{q,1}}\left(\omega\right) \\
		S^{\text{(th)}}_{{q,2}}\left(\omega\right)
	\end{bmatrix},
\end{equation}
where $S^{\text{(th)}}_{{q,1}}(\omega)=4 \gamma_{01} m_{1} k_{\text{B}} T_0	/ \pi$ and $S^{\text{(th)}}_{{q,2}}(\omega)=4 \gamma_{02} m_{2} k_{\text{B}} T_0	/ \pi$ are the PSDs of the thermal forces acting on the first and second particle. The definitions of $\chi_{q,11}, \chi_{q,22}, \chi_{q,12}$ and $\chi_{q,21}$ can be found in Eqs.~\ref{eq:transfer_x11}$-$\ref{eq:transfer_z12}. Figures~\ref{app_fig:fig_2_theory_psd}a and b show calculated PSDs of the directly and sympathetically cooled particles. As the feedback strength is increased, the modal peaks become shorter and broader, merging into a single peak. The broadening is explained by the increasing modal damping rate shown in Fig.~\ref{app_fig:fig_1_modes}b. The merge is due to the combination of two effects: the increase of the modal damping and the shift in the modal frequencies (see Figs.~\ref{app_fig:fig_1_modes}a and b). 

After the merge, as the feedback strength is further increased, the single peaks in Figs.~\ref{app_fig:fig_2_theory_psd}a and b both narrow. In Fig.~\ref{app_fig:fig_2_theory_psd}a, the height of this peak---which corresponds to the motion of the directly cooled particle---continues to decrease. In contrast, in Fig.~\ref{app_fig:fig_2_theory_psd}b, the height of the peak corresponding to the sympathetically cooled particle increases. The narrowing of the peaks and the changes in their heights are consistent with the evolution of the damping rates shown in Fig.~\ref{app_fig:fig_1_modes}b and the change in the modal energy splitting between the particles shown in Figs.~\ref{app_fig:fig_1_modes}c and d. 
From the theoretical PSDs, we calculate the CoM temperatures as a function of the feedback gain~\cite{hebestreit2018calibration}. These calculated temperatures are plotted in Fig.~3d of the main text.

Next, we analyze the optimum cooling rate and the minimum temperature of the sympathetically cooled particle as a function of the parameter $j/2\pi \omega_{0x}$, which characterizes the normal-mode splitting. The optimum feedback cooling rate is defined as the rate at which the minimum temperature is reached. Fig.~\ref{app_fig:fig_2_theory_psd}c shows that the optimum cooling rate increases linearly with the coupling strength. For our system parameters, the slope is 0.7. As a result of the stronger cooling rate, the minimum temperature of the sympathetically cooled particle decreases as $1/\gamma^\mathrm{opt}_{\mathrm{fb},x}$.

\section{Charging particles}
We observe that the charge on trapped particles increases if a pulsed laser hits a metal surface inside the vacuum chamber near the trap. We speculate that the laser desorbs ions from the surface, and that the ions then propagate to the particle and charge it. In the current experiment, we already use a pulsed laser to load particles into our trap via laser-induced acoustic desorption; this laser is focused onto a metal foil on which particles have been evaporated~\cite{bykov2019}. After one or more nanoparticles have been loaded into the trap, for the purposes of charging, we set the laser's pulse energy below the threshold at which nanoparticles are desorbed from the foil. We observe that for particles \SI{300}{\nano\meter} in diameter, the charging process saturates at $\sim \SI{900}{\elementarycharge}$. Thus, for two particles, if both are charged to the saturation level, we achieve an equal number of charges on each one.

\section{Determination of the particles' relative masses and charge}\label{app:size}
\begin{figure}[H]
	\centering
	\includegraphics[width=.7\linewidth]{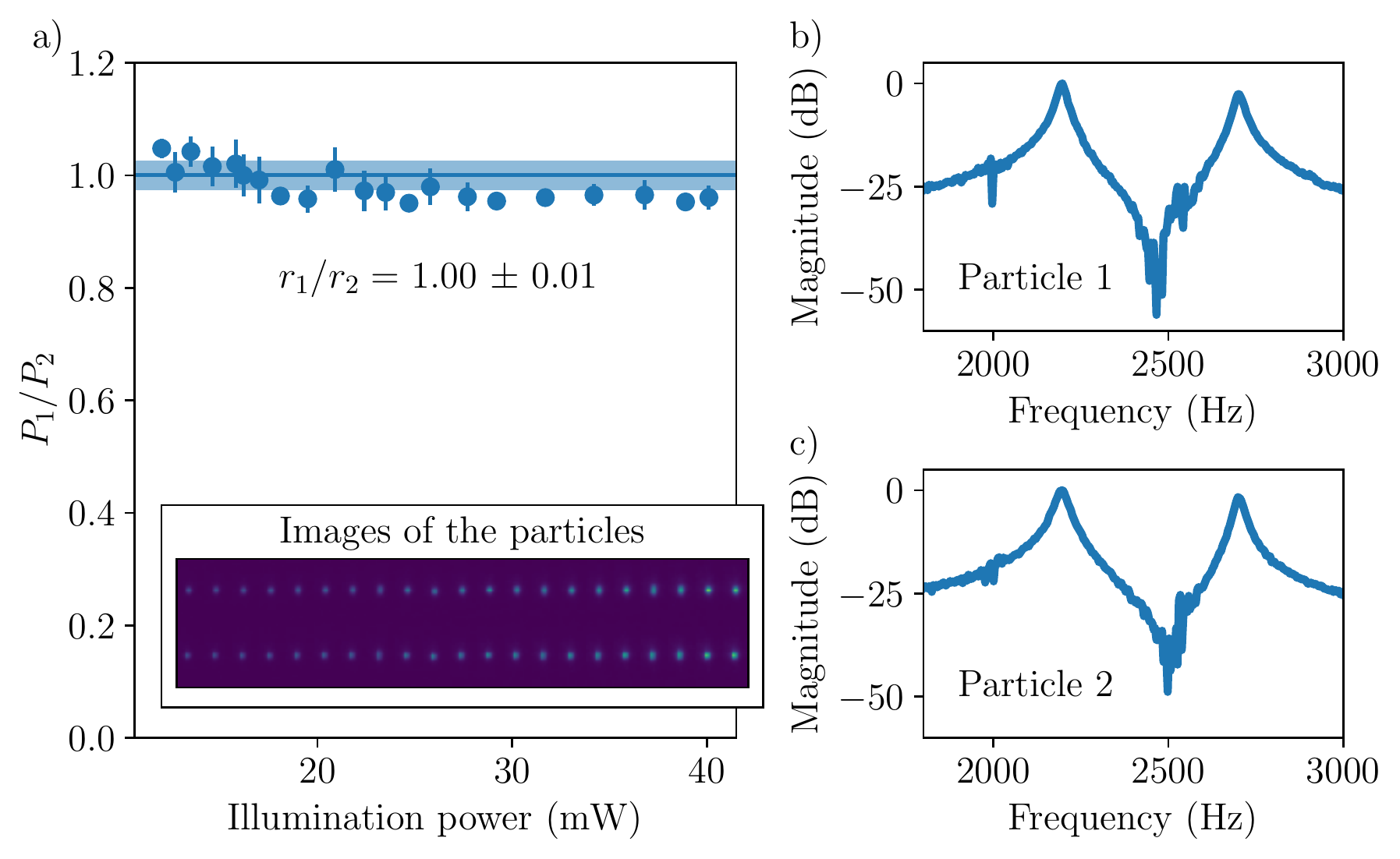}
	\caption{(a) Ratio $P_1/P_2$ of the power scattered by each of the two particles for a range of illumination powers. The line indicates the mean value of the power ratio, and the shaded region indicates the uncertainty. Inset: camera images of the particles at different illumination powers. The power ratio allows the ratio $r_1/r_2$ of the particle radii to be determined. (b) Amplitude response of the first particle as a function of the frequency of an external electric field (in addition to the Paul trap potential). (c) Amplitude response of the second particle as a function of the external field frequency.}
	\label{app_fig:fig_3_characterisation}
\end{figure}
In this section, we first determine the particles' relative masses, for which we measure the power $P_1$ and $P_2$ of the light scattered from each particle. The power is proportional to the sixth power of the particle radius; the ratio of the radii $r_1$ and $r_2$ is thus $r_1 / r_2 = \left(P_1 / P_2\right)^{1/6}$. In Fig.~\ref{app_fig:fig_3_characterisation}, the ratio $P_1/P_2$ is plotted for a range of illumination powers. The scattered power is determined from a camera image: the values of the pixels' brightness are summed in a region of interest around each particle. From these ratios, we find $r_1 / r_2 = 1.00 \pm 0.01$. The inset in Fig.~\ref{app_fig:fig_3_characterisation} shows examples of the images of two particles for different illumination powers.  
 
Next, we determine the particles' relative charge, for which we measure the particles' responses to an external sinusoidal electric force, plotted in Figs.~\ref{app_fig:fig_3_characterisation}b and c. Only resonances corresponding to the particles' in-phase motion are observed. The particles' motion is coupled to the applied electric field through the particles' charge. The driving electric field acts on both particles in phase. For equal charge on the particles, the force will only couple to the in-phase mode, which is consistent with our observation. 
If the charge on the two particles were different, the out-of-phase modes at $\SI{2.0}{}$ and $\SI{2.5}{\kilo\hertz}$ would also be excited. 
A calculation of the response function shows that for the experimental parameters used to obtain the data in Figs.~\ref{app_fig:fig_3_characterisation}b and c, a 5\%  difference in charge number would result in an excitation of the out-of-phase mode with a peak height of 3\% of the in-phase mode, or \SI{-15}{\decibel} for the data shown in Figs.~\ref{app_fig:fig_3_characterisation}b and c. We conclude that the charge on the two particles is equal within 5\%.

\section{Transfer function for sympathetic detection}
\begin{figure}[t]
	\centering
	\includegraphics[width=.9\linewidth]{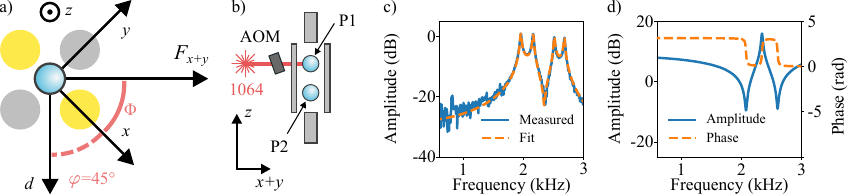}
	\caption{Transfer function for sympathetic detection. (a) Coordinate system, with directions of both force and detection indicated. Cross-sections of the Paul trap RF electrodes are indicated as yellow and gray circles. (b) Schematic of the sympathetic detection experiment. An intensity-modulated \SI{1064}{\nano\meter} laser beam exerts a force on the top particle, P1. Each particle's motion is measured with an independent detection system, as described in the main text. (c) Measured (solid blue line) and fitted (dashed orange line) response of the particle P2 to the drive of the particle P1. (d) Calculated amplitude and phase transfer functions that connect the motion of the particle P1 and P2.}
	\label{app_fig:fig_4_transfer}
\end{figure}
This section provides details on the reconstruction of particle P1's response to an external sinusoidal force, applied only to P1, from a measurement of particle P2's motion. We focus on motion along the radial axes $x$ and $y$; the case of motion along $z$ is analogous. In the final experiment described in the main text, the force is applied to P1 in the direction $\hat{x}+\hat{y}$, and the particles' motion is detected in the direction $\hat{d}=\hat{x}-\hat{y}$, as shown in Figs.~\ref{app_fig:fig_4_transfer}a and b. The force is provided by the \SI{1064}{\nano\meter} laser beam modulated by an acousto-optic modulator, as illustrated in Fig.~\ref{app_fig:fig_4_transfer}b.
According to Eq.~\ref{eq:motion_steady}, the motion of P2 along $\hat{d}$ is described by the expression
\begin{equation}
	x_2\left(\omega\right)  - y_2\left(\omega\right) = \left(\cos\left(\Phi\right)\chi_{x,21}\left(\omega\right) - \sin\left(\Phi\right)\chi_{y,21}\left(\omega\right)\right)F_{1}\left(\omega\right)
\end{equation}
where $\Phi$ is the angle between the applied force direction and the $x$ axis of the trap and $F_{1}\left(\omega\right) = \alpha U_{\text{c}}\left(\omega\right)$. Here, the voltage $U_{\text{c}}$ is the amplitude of the modulation signal applied to the AOM and is generated by a field programmable gate array. The proportionality coefficient $\alpha$ can be considered an unknown parameter. Interferometric detection provides a voltage signal $U_{\text{det}}$ proportional to the particle displacement, which can be written as
\begin{equation}
	\begin{split}
	U_{\text{det}}\left(\omega\right) = \beta\left(x_2\left(\omega\right)  - y_2\left(\omega\right)\right) &= \beta \left(\cos\left(\Phi\right)\chi_{x,21}\left(\omega\right) - \sin\left(\Phi\right)\chi_{y,21}\left(\omega\right)\right) F_{1}\\
	 &= \beta \left(\cos\left(\Phi\right)\chi_{x,21}\left(\omega\right) - \sin\left(\Phi\right)\chi_{y,21}\left(\omega\right)\right) \alpha U_{\text{c}}\left(\omega\right) 
	\end{split}
\end{equation}
for a second unknown parameter $\beta$. The response function can then be expressed as
\begin{equation}\label{eq:response_func}
	\frac{U_{\text{det}}\left(\omega\right)}{U_{\text{c}}\left(\omega\right)} = \alpha \beta \left(\cos\left(\Phi\right)\chi_{x,21}\left(\omega\right) - \sin\left(\Phi\right)\chi_{y,21}\left(\omega\right)\right) = A \left(\cos\left(\Phi\right)\chi_{x,21}\left(\omega\right) - \sin\left(\Phi\right)\chi_{y,21}\left(\omega\right)\right)m = \Psi_2\left(\omega\right),
\end{equation}
where we have combined $\alpha$, $\beta$ and $1/m$ into a single constant $A$. We assume that the particles have identical size, as discussed in Appendix~\ref{app:size}, and as a result, the same mass $m$ and gas damping coefficient $\gamma_0$. 

Figure~\ref{app_fig:fig_4_transfer}c shows the measured amplitude response function $|\Psi_2\left(\omega\right)|$. By fitting this function with Eq.~\ref{eq:response_func}, we extract the parameters $\{\omega_{0x,1}, \omega_{0x,2}, \omega_{0y,1}, \omega_{0y,2}, j_{x}, j_{y}, \gamma_{0}, A,\Phi\}$. Such a fit is also plotted in  Fig.~\ref{app_fig:fig_4_transfer}c. Using Eq.~\ref{eq:motion_steady}, we then obtain the following relationship between the motion of P1 and P2 along the detection direction:
\begin{equation}\label{eq:transfer_function}
	x_1(\omega)  - y_1(\omega) = \left(x_2(\omega)  - y_2(\omega)\right)\frac{\cos\left(\Phi\right)\chi_{x,11} - \sin\left(\Phi\right)\chi_{y,11}}{\cos\left(\Phi\right)\chi_{x,21} - \sin\left(\Phi\right)\chi_{y,21}} = \left(x_2(\omega)  - y_2(\omega)\right) \Theta\left(\omega\right).
\end{equation}
From Eqs.~\ref{eq:transfer_x11}, \ref{eq:transfer_y11}, \ref{eq:transfer_x21} and \ref{eq:transfer_y21} and the fit parameters, we can fully determine the transfer function $\Theta$. The amplitude and phase components of this function are plotted in Fig.~\ref{app_fig:fig_4_transfer}d. Note, however, that in order to compare the detected motion of the particles, as in Fig.~4 of the main text, we need to normalize the detection sensitivity for each particle. To that end, we compare the thermally driven oscillations of the particles at half of the frequency of the slowest mode. We then include this normalization factor in the transfer function $\Theta$.

\bibliography{biblio_sympathetic}